\documentclass{INTERSPEECH2023}

\interspeechcameraready

\title{Adapting an Unadaptable ASR System}
\name{Rao Ma$^\ast$, Mengjie Qian$^\ast$, Mark J. F. Gales, Kate M. Knill\thanks{This paper reports on research supported by Cambridge University Press \& Assessment, a department of The Chancellor, Masters, and Scholars of the University of Cambridge. Mengjie Qian is supported by EPSRC Project EP/V006223/1 (Multimodal Video Search by Examples).}}
\address{ALTA Institute, Machine Intelligence Lab, Department of Engineering, Cambridge University, UK}
\email{\{rm2114,mq227,mjfg100,kmk1001\}@cam.ac.uk}

\usepackage{multirow, multicol}
\usepackage{algorithm, algorithmic, tabularx, graphicx, subfigure, caption}

\newcommand\blfootnote[1]{%
  \begingroup
  \renewcommand\thefootnote{}\footnote{#1}%
  \addtocounter{footnote}{-1}%
  \endgroup
}

\begin{document}
\maketitle
 
\begin{abstract}
As speech recognition model sizes and training data requirements grow, it is increasingly common for systems to only be available via APIs from online service providers rather than having direct access to models themselves. In this scenario it is challenging to adapt systems to a specific target domain. To address this problem we consider the recently released OpenAI Whisper ASR as an example of a large-scale ASR system to assess adaptation methods. An error correction based approach is adopted, as this does not require access to the model, but can be trained from either 1-best or N-best outputs that are normally available via the ASR API. LibriSpeech is used as the primary target domain for adaptation. The generalization ability of the system in two distinct dimensions are then evaluated. First, whether the form of correction model is portable to other speech recognition domains, and secondly whether it can be used for ASR models having a different architecture.

\end{abstract}


\blfootnote{$^\ast$ Equal Contribution.}

\section{Introduction}
Automatic speech recognition (ASR) refers to the task of transcribing the content of speech into readable text. 
The dominant model structure and training schemes have changed over time. 
Early works combined separately trained modules in the decoding process, while recent models employing end-to-end (E2E) architectures achieve start-of-the-art performance \cite{graves2012sequence, chan2016listen}. 
Large-scale models pre-trained on a huge amount of speech data have been released recently.
By training a Transformer based LAS system with more than 680,000 hours of labeled data, Whisper~\cite{radford2022robust} shows accuracy comparable to human transcribers. Google USM~\cite{zhang2023google} learns from 12 million hours speech data in an unsupervised fashion and achieves state-of-the-art performance on multilingual speech recognition tasks.


Although achieving a low word error rate (WER) on test sets is one aim of an ASR system, a useable and reliable system goes beyond this. Other components such as language identification \cite{lopez2014automatic}, speaker diarization \cite{anguera2012speaker}, and text formatting \cite{sunkara2021neural} are all crucial parts of the system and can have a large influence on a client's experience. For personal users or small businesses, building such a pipeline can be arduous, which requires expertise, the acquisition of high-quality speech data, and the availability of computing resources. Therefore, calling APIs from cloud service providers rather than building the entire system from scratch is a more reasonable and economical choice.

With this continuing trend of invoking online ASR services \cite{jeffs2018ok, fortune2022cagr} comes the important question of how to achieve effective customization. Although commercial ASR systems are trained on large-scale data and achieve good performance in general, they are not well-adapted to user-specific needs.
When the incoming speech is from an unseen domain, the transcription tends to be erroneous. Fine-tuning has proven to be an effective method to improve the performance of a pre-trained model. The code and models, however, are the private property of the company, giving users no access in most cases. Further, the cost of fine-tuning an ASR model with billions of parameters makes this option infeasible in practice. 
For example, the largest version of Whisper has 1550M parameters, and Google USM contains 32 Conformer layers with 2B parameters. 

Adaptation of a black-box ASR system is of great interest but there are few prior works. \cite{khandelwal2020black} builds a local accent-tuned ASR model and uses the output from cloud service to guide the decoding of this local model for accent adaptation. 
When the original ASR model is inaccessible, language model (LM) rescoring is another solution. 
\cite{corona2017improving} trains a language model together with the semantic parser to perform reranking on the N-best list from the black-box ASR model. 
The implementation of this approach is subject to the acquisition of N-best hypotheses from service providers.
Recent progress shows that E2E models learn an internal language model (ILM) on the training data, making plain shallow fusion less effective~\cite{meng2021internal}. Different methods for reducing the effect of ILMs have been proposed~\cite{mcdermott2019density, liu22j_interspeech, zeineldeen2021investigating}. Since the implementation of such methods still needs to modify the code running at inference time, they are out of the discussion in this paper. 

Alternatively, post-processing methods such as error detection or correction are promising approaches to adopt \cite{errattahi2018automatic}. The task of error detection is to highlight words within the ASR output that contain mistakes. Error correction also takes the ASR output as input while the aim is to generate a rewritten transcription with lower WER, thereby achieving adaptation when the original model underperforms. Early works designed rule-based systems that are based on statistical analysis~\cite{cucu2013statistical}. Later on, end-to-end models with attention modules have been built which can automatically identify errors considering the context within the sentence and learn the mapping from wrong words to the correct counterparts implicitly~\cite{ren2019fastspeech, guo2019spelling}. With the advent of pre-trained language models (PLM), recent works propose methods to transfer knowledge from PLMs for more accurate detection and correction~\cite{hrinchuk2020correction, zhao21_interspeech, shen-etal-2022-mask}.



In this paper, we take the Whisper model embedded in the OpenAI service as an example to discuss this real-life scenario where we intend to adapt an unadaptable black-box ASR model. 
In the most general case when only an ASR transcription is available, we can get 7.4\% WER reduction over strong baseline results by training an error correction model in the designated domain. With the addition of an N-best list, the performance gain is increased to 15.4\%. Once we train a correction model, it can be applied to utterances from other test sets or from another ASR system in the zero-shot transfer setting. 
By constraining the decoding space in the generation, it shows 6.4\% improvement on out-of-domain test sets over the ASR baselines.

\section{Adaptation Methods}


\subsection{Error Correction Model and Variants}
When the ASR model is not well adapted to the target domain speech, the transcription tends to contain many errors. By building an error correction model on top to correct mistakes in the target domain, we can achieve effective adaptation. A standard error correction model adopts an E2E structure, taking the ASR transcription as input, and is trained to generate the corrected sentence~\cite{guo2019spelling, hrinchuk2020correction, zhao21_interspeech}. To train an error correction model, we do not need access to the ASR system since only decoded hypotheses are needed. Therefore, this method is applicable to adapting a black-box, cloud-based speech-to-text system.
\begin{figure}[!htbp]
    \centering
    \includegraphics[width=5.7cm]{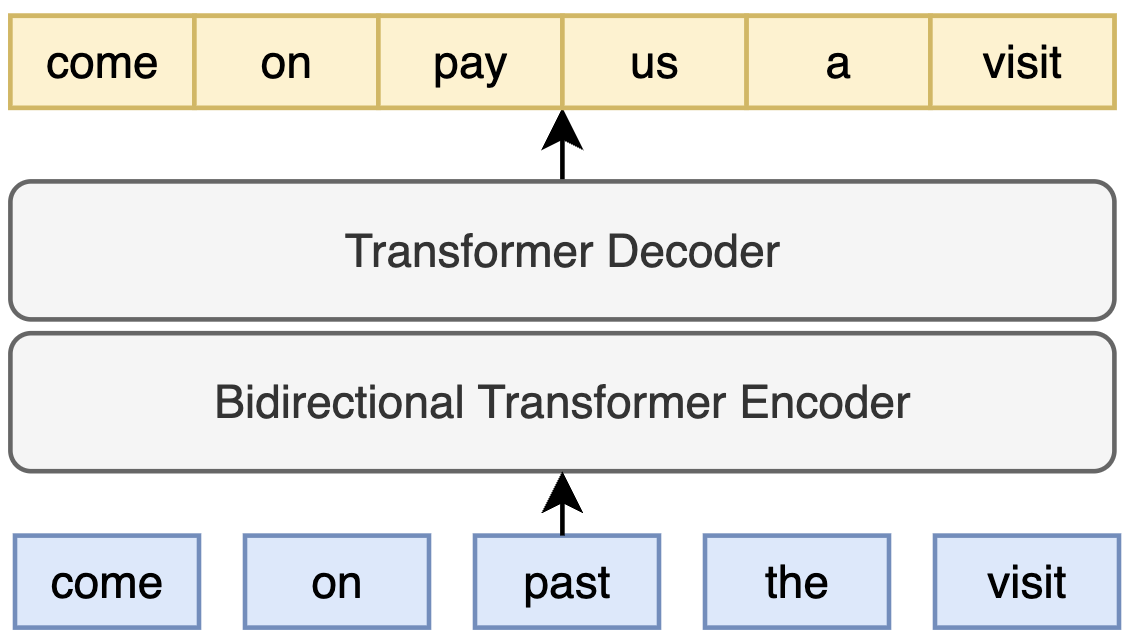}
    \caption{Error correction model structure.}
    \label{fig:correction}
    \vspace{-1mm}
\end{figure}

In addition to text transcriptions, other outputs such as word confidence scores and N-best decoding hypotheses can be returned by some ASR service providers when specified. 
These features can be passed to the error correction model to build a more robust system.
\cite{gekhman-etal-2022-red} incorporates word-level ASR confidence scores into an error detection model, thereby providing the model with information about where the ASR transcript might be wrong. Here, we extend this idea to the error correction task. We follow the practice of using average softmax probability calculated by the ASR model as an estimate of the confidence score for each word. The scores are first quantized into different bins and mapped into confidence embeddings, which are then added to the token embeddings in the input layer. 


\begin{figure}[!htbp]
    \centering
    \subfigure[Extended input with ASR confidence score embedding.]{\includegraphics[width=7.5cm]{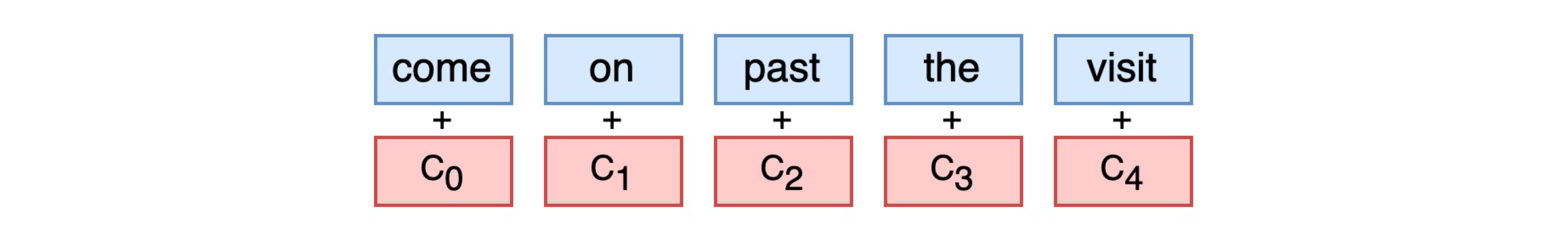}}
    \subfigure[Extended input with phone sequence.]{\includegraphics[width=8cm]{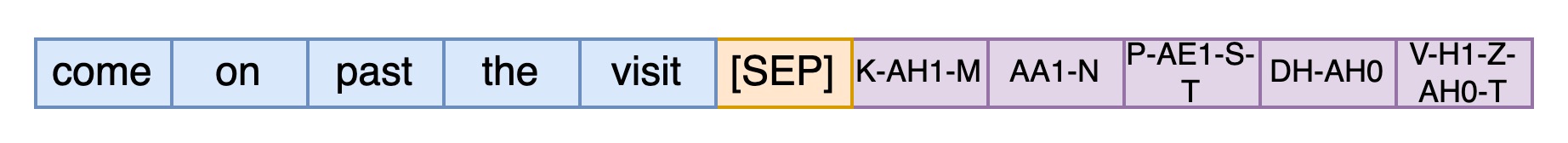}}
    \subfigure[Extended input with ASR N-best hypotheses.]
    {\includegraphics[width=8cm]{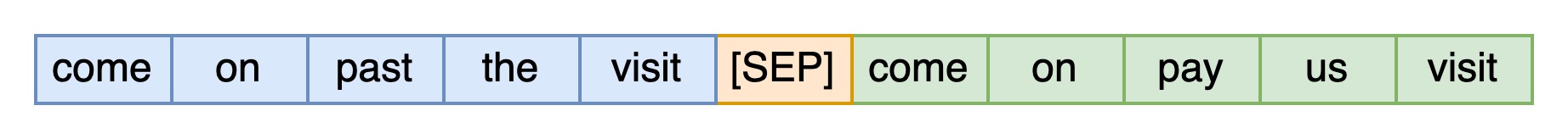}\label{nbest}}
    \vspace{-2mm}
    \caption{Error correction model with additional features.}
    \vspace{-4mm}
    \label{fig:extension}
\end{figure}


In \cite{dutta2022error}, the ASR transcription is converted into a phone sequence and appended in the model input. With this approach, we provide explicit information about the pronunciation of each word, making it easier for the model to correct errors into similar-sounding ones. In addition to the 1-best hypothesis, N-best ASR hypotheses are generated by the ASR system as a byproduct of the beam search process. N-best T5 that leverages the N-best list as input shows significant performance gain over the original model in error correction \cite{ma2023n}. The rationale behind this is that the N-best list contains alternative sequences that are highly possible to be the correct transcription, thus giving the error correction model some cues when making predictions. In the modified input, sentences in the N-best list are sorted with ASR score and concatenated with a special token.


\subsection{Unconstrained vs. N-best Constrained Decoding}
Once trained on the ASR output, the error correction model can be decoded with beam search at inference. Denote the ASR transcription for utterance $x$ as $y$, the target of the error correction model is to find $\hat z =\arg\max_{z} \log P(z|y;\theta_\text{EC})$, where $\hat z$ is the corrected hypothesis. When the N-best hypotheses are available, we can train a correction model with extended input as in Figure \ref{nbest}. Suppose the ASR N-best list is $\mathcal{Y}=\{y_1, ..., y_n\}$, the decoding criterion can be formulated as $\hat z =\arg\max_{z} \log P(z|\mathcal{Y};\theta_\text{EC})$. Since we try to search from all the possible text sequences in the entire decoding space and add no explicit constraints, these methods are denoted as unconstrained decoding in this paper.


To realize more controllable decoding in the generation process, \cite{ma2023n} proposes N-best constrained and lattice constrained decoding methods. The first approach can be directly applied to the black-box adaptation scenario without modifying the ASR model structure. For each hypothesis in the N-best list, scores calculated by the ASR model and the error correction model are linearly combined with an interpolation weight $\lambda$, and the combined score is used for list re-ranking. The modified decoding criterion for the model with N-best list as input is therefore
\begin{equation}
\begin{split}
    \hat z = & \mathop{\arg\max}_{z\in \mathcal{Y}} [(1-\lambda) \cdot \log P(z|x;\theta_\text{ASR}) \\ 
    & + \lambda \cdot \log P(z|\mathcal{Y};\theta_\text{EC})]
\end{split}
\end{equation}
In this way, we constrain the output of the error correction model to appear in the N-best list and we also utilize the ASR score to bias towards sequence with higher acoustic possibility.


\section{Experiments}
\subsection{Setup}
Various sizes of Whisper systems have been released. As the small.en model performs close to the larger models on many datasets and runs considerably faster, it is used as the base ASR system in this paper. The original decoding result only returns the 1-best hypothesis and the sentence-level confidence score for each utterance. We made several changes to the code to actively save the generated N-best lists at inference and to obtain the token-level softmax probabilities to calculate word-level ASR confidence score. This is to simulate the scenario where the ASR service provides extra information for downstream tasks. The code will be made public at a later stage.

We conduct experiments on the public LibriSpeech dataset~\cite{panayotov2015librispeech} to demonstrate the effectiveness of adaptation. The training corpus contains around 960hr speech collected from audio-book reading. The recognition results are evaluated on both ``clean'' and ``other'' subsets of the corpus, each containing over 5hr speech data. At inference, there is an option for Whisper to control whether to decode with or without timestamps. Since time information is not needed in our approach and makes it complicated to get the N-best list, the model is decoded without generating such tokens. Furthermore, Whisper is trained to output transcriptions with punctuation, which makes sentences in the N-best list similar to each other. To obtain the N-best hypotheses with larger context diversity, a list of most common punctuation are suppressed in decoding. In the evaluation, we run text normalization scripts on both reference and hypothesis before calculating WER results following \cite{radford2022robust}.



For the error correction task, we transfer knowledge from PLM with standard E2E structures. Pre-trained T5-base~\cite{raffel2020exploring} and BART-base~\cite{lewis2020bart} models are utilized to initialize the model parameters respectively. We first run the Whisper model on the training set of LibriSpeech to collect the decoding results of 960hr speech data. The 1-best decoding results with ASR errors and paired reference text are then used to train the error correction model for 3 epochs. In fine-tuning, the Adam scheduler is used with an initial learning rate of 5e-5. 

Language model rescoring, another commonly used adaptation method is tested in our experiments.
We train two language models with different structures. A GPT-2-base model \cite{radford2019language} is fine-tuned on the reference text of the 960hr training speech, which consists of 281K sentences. The model contains 12-layer Transformer blocks and is trained with a learning rate of 5e-5 for 5 epochs. Since LibriSpeech also provides a large text corpus containing 40.7M sentences, another big LSTMLM is trained to examine if Whisper benefits from usage of additional text data. This model has 4 LSTM layers with a hidden size of 2048 and the training process follows the ESPnet~\cite{watanabe2018espnet} recipe. The average word-level perplexity for GPT-2 and LSTMLM on the reference text of test sets are 118 and 83 respectively.

\subsection{Adaptation via Error Correction}

Due to the effectiveness of large-scale pre-training and multitask learning, Whisper shows incredibly good performance of 3.52\% and 7.37\% on the LibriSpeech test sets. 
We compare different adaptation strategies using LM rescoring and error correction in Table \ref{tab:whisper_t5_rescore_nonorm_constr}. 
By performing 10-best list rescoring with LSTMLM or a finetuned GPT-2 model, the performance on test\_other is marginally improved while the WER on the clean subset is worsened. This suggests that external LMs have little impact on the results when Whisper has already learned a strong implicit language model on large-scale training data.

Despite this, the T5 error correction model achieves a performance gain of 10.2\% and 4.6\% on the test\_clean and test\_other sets. A model fine-tuned from BART also improves over the ASR baseline but performs slightly worse than T5 model. These results show the promising potential of using error correction models for domain adaptation even when only the 1-best transcription is returned from the black-box system.

\begin{table}[htbp!]
\caption{Comparison of LM rescoring and error correction.}
\vspace{-1mm}
    \label{tab:whisper_t5_rescore_nonorm_constr}
    \centering
    \begin{tabular}{l|l|c|c|c|c}
        \toprule
        \multirow{2}*{Model} & \multirow{2}*{Method} & \multicolumn{2}{c|}{Dev} & \multicolumn{2}{c}{Test} \\
        & & clean & other & clean & other \\
        \midrule
        Baseline & - & 3.62 & 6.61 & 3.52 & 7.37 \\
        \midrule
        GPT-2 & Rescoring & 3.96 & 6.53 & 4.17 & 7.38 \\
        LSTMLM & Rescoring & 3.92 & 6.49 & 4.11 & 7.23 \\
        \midrule
        BART & Correction & 3.25 & 6.21 & 3.34 & 7.07 \\
        T5 & Correction & \textbf{3.03} & \textbf{6.08} & \textbf{3.16} & \textbf{7.03}\\
        \bottomrule
    \end{tabular}
    \vspace{-3mm}
\end{table}

\begin{figure}[!htbp]
    \centering
    \vspace{1mm}
    \includegraphics[width=6cm]{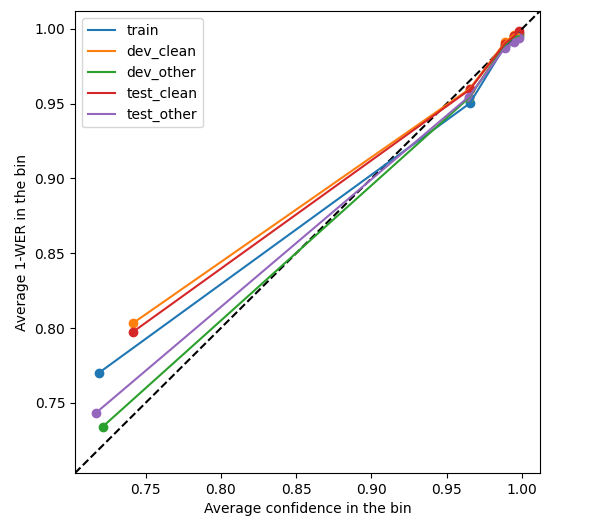}
    \caption{Confidence score analysis on LibriSpeech.}
    \label{fig:robart_exp}
    \vspace{-1mm}
\end{figure}

In Table \ref{tab:t5_conf_phone} we show the results of using additional features in the T5 error correction model. When adding confidence score embeddings to the T5 encoder, slightly better performance can be observed. Here the results are obtained using 5 bins to quantize confidence scores. Increasing the bin size yields similar performance. Figure \ref{fig:robart_exp} plots the average confidence score and average $1-$WER for words in each quantized bin, which shows that the softmax probabilities of Whisper are well-calibrated. According to Table \ref{tab:t5_conf_phone}, appending the phone sequence to the input transcription leads to degraded performance. These results suggest that ASR confidence scores and phone sequence features are not necessary for training an error correction model.



By taking ASR N-best hypotheses as input, more useful information can be leveraged by the encoder. When we increase the 1-best input to the 10-best ASR outputs, a WER reduction of 15.4\% compared to the baseline can be seen on the test sets. Therefore, using N-best lists as input can effectively improve the performance of the downstream error correction task. For the model trained with 10-best input, N-best constrained decoding results are also given, showing a WER reduction of 10.5\% compared to the Whisper baseline. Since the error correction model is trained on 960hr training speech, it learns well to detect and correct errors for utterances from the LibriSpeech corpus. With constrained decoding, the model is restricted to not generating text outside the N-best list, which explains the degradation compared to the unconstrained decoding mode.

\begin{table}[htbp!]
    \caption{Error correction results using various input features.}
    \vspace{-2mm}
    \label{tab:t5_conf_phone}
    \centering
    \begin{tabular}{ll|c|c|c|c}
        \toprule
        \multicolumn{2}{l|}{\multirow{2}*{Model}} & \multicolumn{2}{c|}{Dev} & \multicolumn{2}{c}{Test} \\
        && clean & other & clean & other \\
        \midrule
        \multicolumn{2}{l|}{1-best} & 3.03 & 6.08 & 3.16 & 7.03 \\
        \multicolumn{2}{l|}{\hspace{5mm} + confidence} & 3.01 & 6.03 & 3.11 & 7.04 \\
        \multicolumn{2}{l|}{\hspace{5mm} + phones} & 3.05 & 6.29 & 3.21 & 7.05 \\ 
        \midrule
        10-best & uncon & \textbf{2.60} & \textbf{5.79} & \textbf{2.90} & \textbf{6.39} \\
        10-best & constr & 2.87 & 5.87 & 3.10 & 6.69 \\
        \bottomrule
    \end{tabular}
\end{table}

Normally, pre-trained ASR systems learn from huge amounts of speech data that effectively cover multiple topics and are from diversified speakers. For example, Whisper is reported to yield low WER on more than 20 speech recognition test sets. With the adaptation process focusing on a specific domain, our results indicate that the recognition accuracy on the target test sets can be noticeably improved. Whether utterances from other datasets show similar performance gains is another intriguing question. In this section, we examine the generalization ability of the adaptation method.
For the results in Table \ref{tab:whisper_t5_other_dataset}, error correction models trained on LibriSpeech transcriptions are directly applied to the ASR output from other test sets without fine-tuning. 
The models are evaluated on three standard test sets from TED-LIUM 3~\cite{hernandez2018ted}, Artie bias corpus~\cite{meyer2020artie}, and MGB3 dev17b~\cite{bell2015mgb}, which are abbreviated into TED, Artie and MGB.


\begin{table}[htbp!]
    \caption{Error correction results on other datasets by models trained on LibriSpeech in the transfer setting.}
    \label{tab:whisper_t5_other_dataset}
    \centering
    \begin{tabular}{ll|cc|ccc}
        \toprule
        \multicolumn{2}{l|}{\multirow{2}*{Model}} & \multicolumn{2}{c|}{LibriSpeech} & \multicolumn{3}{c}{Other sets} \\
        & & clean & other & TED & Artie & MGB \\
        \midrule
        \multicolumn{2}{l|}{ASR Baseline} & 3.52 & 7.37 & 3.89 & 9.03 & 13.10 \\
        \midrule
        \multirow{1}*{1-best} & uncon & 3.16 & 7.03 & 4.78 & 9.27 & 17.29 \\
        \multirow{1}*{10-best} & uncon & \textbf{2.90} & \textbf{6.39} & 4.56 & 9.16 & 22.88 \\
        \multirow{1}*{10-best} & constr & 3.10 & 6.69 & \textbf{3.64} & \textbf{8.14} & \textbf{12.71}\\
        \bottomrule
    \end{tabular}
\end{table}

With unconstrained decoding, the correction model generates worse output compared to the ASR baseline due to the mismatch of training and evaluation corpora. Since the T5 model is fine-tuned on LibriSpeech data, ILM in the decoder characterizes sentences from LibriSpeech \cite{meng2021internal}. When we apply N-best constrained decoding instead, an average of 6.4\% WER reduction can be observed on the three out-of-domain test sets. 
Results suggest that the knowledge learned in the model to correct homophones is not limited to utterances within the adapted domain but generalizes well. 
Therefore, for the incoming speech from different sources, we can use one unified error correction model. For utterances that match the trained domain, unconstrained decoding works better, whereas decoding constraints are needed for the model to perform well on other test sets.



\begin{table}[htbp!]
    \caption{Error correction results for Conformer-Transducer ASR model on LibriSpeech in the transfer setting.}
    \label{tab:whisper_t5_rescore_ct}
    \centering
    \begin{tabular}{ll|cc|cc}
        \toprule
        \multicolumn{2}{l|}{\multirow{2}*{Model}} & \multicolumn{2}{c|}{Dev} & \multicolumn{2}{c}{Test} \\
        && clean & other & clean & other \\
        \midrule
        \multicolumn{2}{l|}{ASR Baseline} & 2.61 & 6.79 & 2.79 & 6.90 \\
        \midrule
        \multirow{1}*{1-best} & uncon & 2.64 & 6.75 & 2.78 & 6.92 \\
        \multirow{1}*{10-best} & uncon & 4.07 & 7.93 & 3.86 & 7.72 \\
        \multirow{1}*{10-best} & constr & \textbf{2.48} & \textbf{6.42} & \textbf{2.62} & \textbf{6.65} \\
        \bottomrule
    \end{tabular}
\end{table}
We also apply the T5 model trained on Whisper decoding results to correct the output from a different ASR model. A novel Conformer-Transducer model containing 12 encoder layers is utilized here. The model is trained on 960hr LibriSpeech data following the ESPnet recipe. As shown in Table \ref{tab:whisper_t5_rescore_ct}, N-best T5 improves over the strong baseline results by $\sim$4.8\% with N-best constrained decoding. 
In line with previous experiments, the model with the 10-best input and 10-best constrained decoding space performs the best in the transfer setting. The results indicate that the trained error correction models enable plug-and-play support for existing ASR systems. When we switch ASR services, the same correction model can be effectively used for target domain adaptation without re-training, which is time-efficient and cost-saving.


\subsection{Why is N-best Input Effective?}





\begin{figure}[htbp!]
    \vspace{-2mm}
    \centering
    \includegraphics[trim=0 0 0 1mm, width=6.7cm]{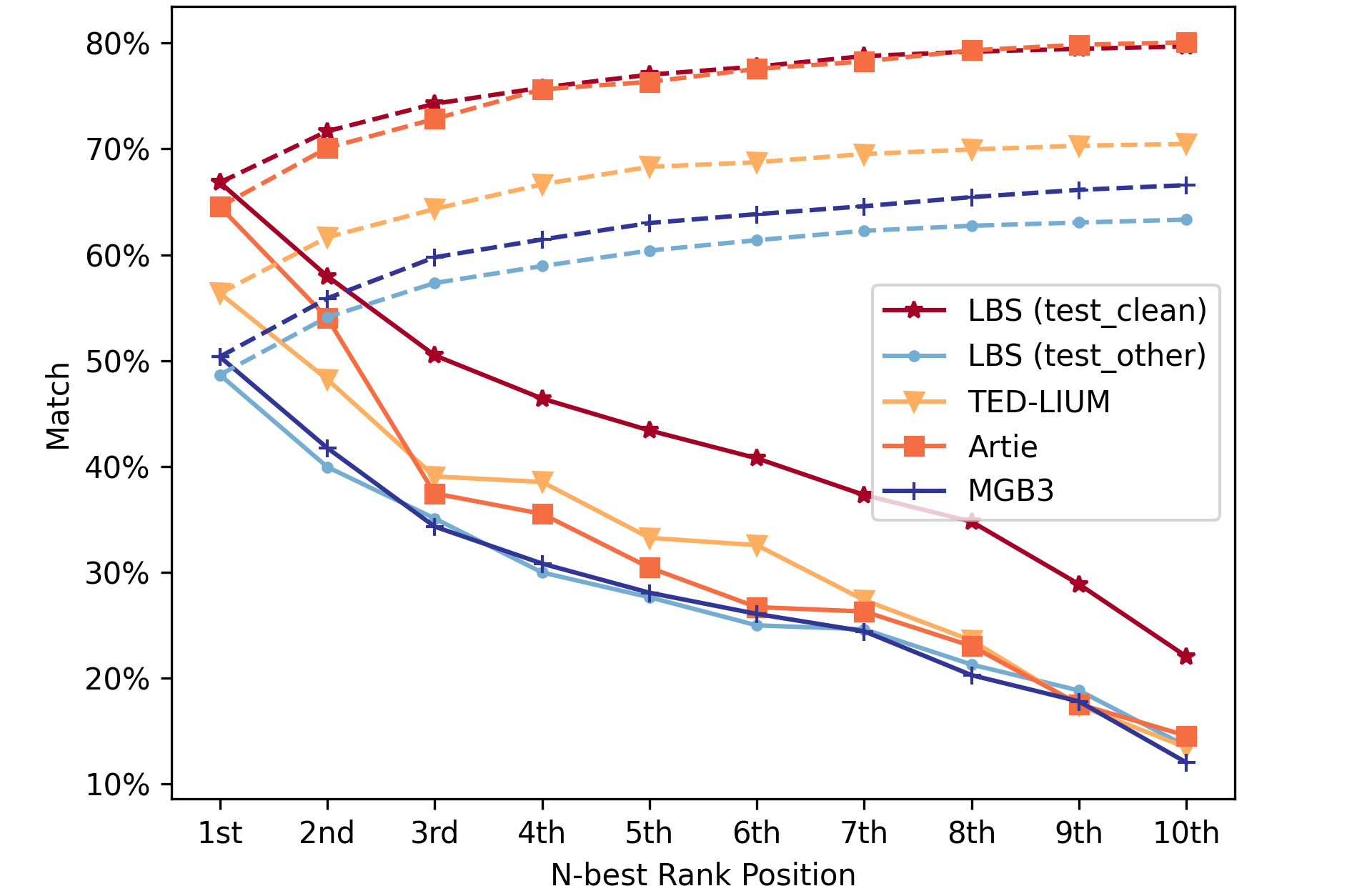}
    \vspace{-2mm}
    \caption{N-best list analysis. Solid lines: Percentage of $n^{th}$ best hypothesis that matches the reference. Dashed lines: Percentage of top $n$ best hypotheses list that contains the reference.}
    \label{fig:nbest_match}
    \vspace{-2mm}
\end{figure}

From previous sections, we can see that adopting an N-best list rather than using the 1-best transcription as input largely improves the error correction performance. In this section, we examine where the performance boost comes from. In the dashed lines of Figure \ref{fig:nbest_match}, we plot the percentage of utterances that contain the reference text in the N-best list with increasing $n$, i.e. the oracle $1-$SER in the top $n$ ASR hypotheses. When using a larger $n$, it is more likely that the reference text appears in the input, reducing the difficulty for the error correction model to recover the correct answer.

\begin{table}[htbp!]
    \caption{Ablation analysis with disturbed 10-best list.}
    \vspace{-1mm}
    \label{tab:whisper_t5_rescore_rnd}
    \centering
    \begin{tabular}{l|cc|ccc}
        \toprule
        \multirow{2}*{10-best} & \multicolumn{2}{c|}{LibriSpeech} & \multicolumn{3}{c}{Other sets} \\
        & clean & other & TED & Artie & MGB \\
        \midrule
        Sorted & \textbf{2.90} & \textbf{6.39} & \textbf{3.64} & \textbf{8.14} & \textbf{12.71} \\
        Randomized & 3.31 & 6.82 & 3.74 & 8.50 &  13.01 \\
        Reversed & 3.50 & 7.18 & 3.75 & 8.57 & 12.99\\
        \bottomrule
    \end{tabular}
    \vspace{-1mm}
\end{table}

Figure \ref{fig:nbest_match} also exhibits that as we increase the rank position, the percentage of $n$-th hypotheses matching the reference decreases, suggesting sentences with lower ASR scores are of worse quality. In the training and evaluation of N-best T5, hypotheses are concatenated in order while we do not add explicit cues about the ranking information. Can the correction model induce such knowledge from the input and use it to achieve better performance? We ran experiments using a randomly shuffled N-best list or a list sorted with ASR scores in reverse order. In Table \ref{tab:whisper_t5_rescore_rnd} we applied unconstrained decoding to the LibriSpeech test sets and used N-best constrained decoding for other sets. The results show that randomizing the N-best list in decoding brings degradation while the model with reversed input hypotheses yields the worst performance. Therefore, ranking information is implicitly learned and preserved and is significant for the N-best T5 model to achieve good performance.



\section{Conclusions}
In this paper we have examined how it is possible to adapt an ASR system to a particular domain without direct access to the system itself. By exploiting 1-best and N-best outputs from the system within an error correction framework it is possible to take the standard output from freely available ASR systems and adapt them to the target domain. We further explore whether error correction modules trained for a specific domain can be applied to other speech domains, as well as other ASR systems.

\bibliographystyle{IEEEtran}
\bibliography{mybib}

\begin{thebibliography}{10}
\providecommand{\url}[1]{#1}
\csname url@samestyle\endcsname
\providecommand{\newblock}{\relax}
\providecommand{\bibinfo}[2]{#2}
\providecommand{\BIBentrySTDinterwordspacing}{\spaceskip=0pt\relax}
\providecommand{\BIBentryALTinterwordstretchfactor}{4}
\providecommand{\BIBentryALTinterwordspacing}{\spaceskip=\fontdimen2\font plus
\BIBentryALTinterwordstretchfactor\fontdimen3\font minus
  \fontdimen4\font\relax}
\providecommand{\BIBforeignlanguage}[2]{{%
\expandafter\ifx\csname l@#1\endcsname\relax
\typeout{** WARNING: IEEEtran.bst: No hyphenation pattern has been}%
\typeout{** loaded for the language `#1'. Using the pattern for}%
\typeout{** the default language instead.}%
\else
\language=\csname l@#1\endcsname
\fi
#2}}
\providecommand{\BIBdecl}{\relax}
\BIBdecl

\bibitem{graves2012sequence}
A.~Graves, ``Sequence transduction with recurrent neural networks,''
  \emph{arXiv preprint arXiv:1211.3711}, 2012.

\bibitem{chan2016listen}
W.~Chan, N.~Jaitly, Q.~Le, and O.~Vinyals, ``Listen, attend and spell: A neural
  network for large vocabulary conversational speech recognition,'' in
  \emph{2016 IEEE international conference on acoustics, speech and signal
  processing (ICASSP)}.\hskip 1em plus 0.5em minus 0.4em\relax IEEE, 2016, pp.
  4960--4964.

\bibitem{radford2022robust}
A.~Radford, J.~W. Kim, T.~Xu, G.~Brockman, C.~McLeavey, and I.~Sutskever,
  ``Robust speech recognition via large-scale weak supervision,'' \emph{arXiv
  preprint arXiv:2212.04356}, 2022.

\bibitem{zhang2023google}
Y.~Zhang, W.~Han, J.~Qin, Y.~Wang, A.~Bapna, Z.~Chen, N.~Chen, B.~Li,
  V.~Axelrod, G.~Wang \emph{et~al.}, ``Google {USM}: Scaling automatic speech
  recognition beyond 100 languages,'' \emph{arXiv preprint arXiv:2303.01037},
  2023.

\bibitem{lopez2014automatic}
I.~Lopez-Moreno, J.~Gonzalez-Dominguez, O.~Plchot, D.~Martinez,
  J.~Gonzalez-Rodriguez, and P.~Moreno, ``Automatic language identification
  using deep neural networks,'' in \emph{2014 IEEE international conference on
  acoustics, speech and signal processing (ICASSP)}.\hskip 1em plus 0.5em minus
  0.4em\relax IEEE, 2014, pp. 5337--5341.

\bibitem{anguera2012speaker}
X.~Anguera, S.~Bozonnet, N.~Evans, C.~Fredouille, G.~Friedland, and O.~Vinyals,
  ``Speaker diarization: A review of recent research,'' \emph{IEEE Transactions
  on audio, speech, and language processing}, vol.~20, no.~2, pp. 356--370,
  2012.

\bibitem{sunkara2021neural}
M.~Sunkara, C.~Shivade, S.~Bodapati, and K.~Kirchhoff, ``Neural inverse text
  normalization,'' in \emph{2021 IEEE International Conference on Acoustics,
  Speech and Signal Processing (ICASSP)}.\hskip 1em plus 0.5em minus
  0.4em\relax IEEE, 2021, pp. 7573--7577.

\bibitem{jeffs2018ok}
C.~Harrison, ``{OK Google, Siri, Alexa, Cortana; Can you tell me some stats on
  voice search},'' \emph{Edit Agency}, 2018.

\bibitem{fortune2022cagr}
B.~Neeley, ``{Speech and voice recognition market size, with 23.7\% CAGR},''
  {https://biz.crast.net/speech-and-voice-recognition-market-size-with-23-7-cagr/},
  November 2022.

\bibitem{khandelwal2020black}
K.~Khandelwal, P.~Jyothi, A.~Awasthi, and S.~Sarawagi, ``Black-box adaptation
  of {ASR} for accented speech,'' in \emph{Proc. Interspeech 2020}, 2020, pp.
  1281--1285.

\bibitem{corona2017improving}
R.~Corona, J.~Thomason, and R.~Mooney, ``Improving black-box speech recognition
  using semantic parsing,'' in \emph{Proceedings of the Eighth International
  Joint Conference on Natural Language Processing (Volume 2: Short Papers)},
  2017, pp. 122--127.

\bibitem{meng2021internal}
Z.~Meng, S.~Parthasarathy, E.~Sun, Y.~Gaur, N.~Kanda, L.~Lu, X.~Chen, R.~Zhao,
  J.~Li, and Y.~Gong, ``Internal language model estimation for domain-adaptive
  end-to-end speech recognition,'' in \emph{2021 IEEE Spoken Language
  Technology Workshop (SLT)}.\hskip 1em plus 0.5em minus 0.4em\relax IEEE,
  2021, pp. 243--250.

\bibitem{mcdermott2019density}
E.~McDermott, H.~Sak, and E.~Variani, ``A density ratio approach to language
  model fusion in end-to-end automatic speech recognition,'' in \emph{2019 IEEE
  Automatic Speech Recognition and Understanding Workshop (ASRU)}.\hskip 1em
  plus 0.5em minus 0.4em\relax IEEE, 2019, pp. 434--441.

\bibitem{liu22j_interspeech}
Y.~Liu, R.~Ma, H.~Xu, Y.~He, Z.~Ma, and W.~Zhang, ``{Internal Language Model
  Estimation Through Explicit Context Vector Learning for Attention-based
  Encoder-decoder ASR},'' in \emph{Proc. Interspeech 2022}, 2022, pp.
  1666--1670.

\bibitem{zeineldeen2021investigating}
M.~Zeineldeen, A.~Glushko, W.~Michel, A.~Zeyer, R.~Schlüter, and H.~Ney,
  ``{Investigating Methods to Improve Language Model Integration for
  Attention-Based Encoder-Decoder ASR Models},'' in \emph{Proc. Interspeech
  2021}, 2021, pp. 2856--2860.

\bibitem{errattahi2018automatic}
R.~Errattahi, A.~El~Hannani, and H.~Ouahmane, ``Automatic speech recognition
  errors detection and correction: A review,'' \emph{Procedia Computer
  Science}, vol. 128, pp. 32--37, 2018.

\bibitem{cucu2013statistical}
H.~Cucu, A.~Buzo, L.~Besacier, and C.~Burileanu, ``Statistical error correction
  methods for domain-specific {ASR} systems,'' in \emph{Statistical Language
  and Speech Processing: First International Conference, SLSP 2013, Tarragona,
  Spain, July 29-31, 2013. Proceedings 1}.\hskip 1em plus 0.5em minus
  0.4em\relax Springer, 2013, pp. 83--92.

\bibitem{ren2019fastspeech}
Y.~Ren, Y.~Ruan, X.~Tan, T.~Qin, S.~Zhao, Z.~Zhao, and T.-Y. Liu, ``Fastspeech:
  Fast, robust and controllable text to speech,'' \emph{Advances in neural
  information processing systems}, vol.~32, 2019.

\bibitem{guo2019spelling}
J.~Guo, T.~N. Sainath, and R.~J. Weiss, ``A spelling correction model for
  end-to-end speech recognition,'' in \emph{2019 IEEE International Conference
  on Acoustics, Speech and Signal Processing (ICASSP)}.\hskip 1em plus 0.5em
  minus 0.4em\relax IEEE, 2019, pp. 5651--5655.

\bibitem{hrinchuk2020correction}
O.~Hrinchuk, M.~Popova, and B.~Ginsburg, ``Correction of automatic speech
  recognition with transformer sequence-to-sequence model,'' in \emph{2020 IEEE
  International Conference on Acoustics, Speech and Signal Processing
  (ICASSP)}.\hskip 1em plus 0.5em minus 0.4em\relax IEEE, 2020, pp. 7074--7078.

\bibitem{zhao21_interspeech}
Y.~Zhao, X.~Yang, J.~Wang, Y.~Gao, C.~Yan, and Y.~Zhou, ``{BART Based Semantic
  Correction for Mandarin Automatic Speech Recognition System},'' in
  \emph{Proc. Interspeech 2021}, 2021, pp. 2017--2021.

\bibitem{shen-etal-2022-mask}
K.~Shen, Y.~Leng, X.~Tan, S.~Tang, Y.~Zhang, W.~Liu, and E.~Lin, ``Mask the
  correct tokens: An embarrassingly simple approach for error correction,'' in
  \emph{Proceedings of the 2022 Conference on Empirical Methods in Natural
  Language Processing}.\hskip 1em plus 0.5em minus 0.4em\relax Association for
  Computational Linguistics, 2022, pp. 10\,367--10\,380.

\bibitem{gekhman-etal-2022-red}
Z.~Gekhman, D.~Zverinski, J.~Mallinson, and G.~Beryozkin, ``{RED}-{ACE}: Robust
  error detection for {ASR} using confidence embeddings,'' in \emph{Proceedings
  of the 2022 Conference on Empirical Methods in Natural Language
  Processing}.\hskip 1em plus 0.5em minus 0.4em\relax Association for
  Computational Linguistics, 2022, pp. 2800--2808.

\bibitem{dutta2022error}
S.~Dutta, S.~Jain, A.~Maheshwari, G.~Ramakrishnan, and P.~Jyothi, ``Error
  correction in {ASR} using sequence-to-sequence models,'' \emph{arXiv preprint
  arXiv:2202.01157}, 2022.

\bibitem{ma2023n}
R.~Ma, M.~J. Gales, K.~Knill, and M.~Qian, ``N-best {T5}: Robust {ASR} error
  correction using multiple input hypotheses and constrained decoding space,''
  \emph{arXiv preprint arXiv:2303.00456}, 2023.

\bibitem{panayotov2015librispeech}
V.~Panayotov, G.~Chen, D.~Povey, and S.~Khudanpur, ``Librispeech: an {ASR}
  corpus based on public domain audio books,'' in \emph{2015 IEEE international
  conference on acoustics, speech and signal processing (ICASSP)}.\hskip 1em
  plus 0.5em minus 0.4em\relax IEEE, 2015, pp. 5206--5210.

\bibitem{raffel2020exploring}
C.~Raffel, N.~Shazeer, A.~Roberts, K.~Lee, S.~Narang, M.~Matena, Y.~Zhou,
  W.~Li, and P.~J. Liu, ``Exploring the limits of transfer learning with a
  unified text-to-text transformer,'' \emph{The Journal of Machine Learning
  Research}, vol.~21, no.~1, pp. 5485--5551, 2020.

\bibitem{lewis2020bart}
M.~Lewis, Y.~Liu, N.~Goyal, M.~Ghazvininejad, A.~Mohamed, O.~Levy, V.~Stoyanov,
  and L.~Zettlemoyer, ``{BART}: Denoising sequence-to-sequence pre-training for
  natural language generation, translation, and comprehension,'' in
  \emph{Proceedings of the 58th Annual Meeting of the Association for
  Computational Linguistics}, 2020, pp. 7871--7880.

\bibitem{radford2019language}
A.~Radford, J.~Wu, R.~Child, D.~Luan, D.~Amodei, I.~Sutskever \emph{et~al.},
  ``Language models are unsupervised multitask learners,'' \emph{OpenAI blog},
  vol.~1, no.~8, p.~9, 2019.

\bibitem{watanabe2018espnet}
S.~Watanabe, T.~Hori, S.~Karita, T.~Hayashi, J.~Nishitoba, Y.~Unno, N.-E.~Y.
  Soplin, J.~Heymann, M.~Wiesner, N.~Chen \emph{et~al.}, ``{ESPnet}: End-to-end
  speech processing toolkit,'' in \emph{Proc. Interspeech 2018}, 2018, pp.
  2207--2211.

\bibitem{hernandez2018ted}
F.~Hernandez, V.~Nguyen, S.~Ghannay, N.~Tomashenko, and Y.~Esteve, ``{TED-LIUM}
  3: Twice as much data and corpus repartition for experiments on speaker
  adaptation,'' in \emph{Speech and Computer: 20th International Conference,
  SPECOM 2018, Leipzig, Germany, September 18--22, 2018, Proceedings 20}.\hskip
  1em plus 0.5em minus 0.4em\relax Springer, 2018, pp. 198--208.

\bibitem{meyer2020artie}
J.~Meyer, L.~Rauchenstein, J.~D. Eisenberg, and N.~Howell, ``Artie bias corpus:
  An open dataset for detecting demographic bias in speech applications,'' in
  \emph{Proceedings of the Twelfth Language Resources and Evaluation
  Conference}, 2020, pp. 6462--6468.

\bibitem{bell2015mgb}
P.~Bell, M.~J. Gales, T.~Hain, J.~Kilgour, P.~Lanchantin, X.~Liu, A.~McParland,
  S.~Renals, O.~Saz, M.~Wester \emph{et~al.}, ``The {MGB} challenge: Evaluating
  multi-genre broadcast media recognition,'' in \emph{2015 IEEE Workshop on
  Automatic Speech Recognition and Understanding (ASRU)}.\hskip 1em plus 0.5em
  minus 0.4em\relax IEEE, 2015, pp. 687--693.

\end{thebibliography}

\end{document}